\RequirePackage[l2tabu, orthodox]{nag}
\documentclass[twocolumn,amsmath,amssymb,aps,showkeys,nofootinbib]{revtex4-2}

\usepackage{graphicx}
\graphicspath{{figures/}}
\usepackage{dcolumn}
\usepackage{bm}
\usepackage[hidelinks]{hyperref}

\usepackage{titlesec}
\titleformat{\section}[hang]{\normalfont\bfseries}{\thesection.}{0.5em}{}[]
\titlespacing{\section}{0pt}{*2}{*1}

\begin{document}

\title{COVID-19 epidemiology as emergent behavior on a dynamic transmission forest}
\author{Niket Thakkar}
\email{niket.thakkar@gatesfoundation.org}
\homepage[\\]{https://github.com/NThakkar-IDM/covid_and_stat_mech}
\author{Mike Famulare}
\affiliation{%
The Institute for Disease Modeling\\
Global Health | Bill $\&$ Melinda Gates Foundation\\
Seattle, Washington 98109
}%
\date{\today}

\begin{abstract}
In this paper we create a compartmental, stochastic process model of SARS-CoV-2 transmission, where the process's mean and variance have distinct dynamics. The model is fit to time series data from Washington from January 2020 to March 2021 using a deterministic, biologically-motivated signal processing approach, and we show that the model's hidden states, like population prevalence, agree with survey and other estimates. Then, in the paper's second half, we demonstrate that the same model can be reframed as a branching process with a dynamic degree distribution. This perspective allows us to generate approximate transmission trees and estimate some higher order statistics, like the clustering of cases as outbreaks, which we find to be consistent with related observations from contact tracing and phylogenetics. 
\end{abstract}
\maketitle

\section{Epidemiology and statistical mechanics}
Basic situational awareness, sometimes even at the level of deciding if trends are rising or falling \cite{cbs2020}, has been a consistent public health challenge during the pandemic. Data are collected and reported at often overwhelming speed and volume, and to make evidence-based decisions, insights from different sources have to be balanced in real-time. 

Part of the difficulty is that the data address the situation on a variety of levels. Individual-level data, like the progression of symptoms or a time series of antibody titer, are separated in scale from population-level data, like a time series of cases or a seroprevalence survey. Meanwhile, findings from outbreak investigations or from genetic sequencing -- data that can only be interpreted for collections of interacting individuals within the population -- live at some level in between. 

The aspiration for situational awareness is understanding how all these pieces fit together in a consistent epidemiology. While it's sometimes conceptually appealing to build from the ground up, first characterizing the biology within individuals and then carefully working towards larger collections, it becomes clear quickly that  ``more is different'' \cite{anderson1972more}. Population dynamics both poorly constrain and are poorly constrained by tractable interaction models \cite{silverman2021situating}, and so it's often neither feasible nor desirable to try to approach the daily practice of public health in this way \cite{oha2020}. In analogy to statistical mechanics, where the major philosophical lesson is that understanding the emergent macroscopic physics (i.e. thermodynamics) requires us to coarsen the microscopic physics (i.e. classical mechanics), good situational awareness requires us to distill the biology at all levels and to better understand how a single model might capture epidemiological features at each scale.

Along those lines, developing intuitive mathematical relationships between observations at the individual and population levels is this paper's main motivation. This is, of course, a very broad goal, and we work towards it here in just two specific instances. 

First, we show that characterizing pathogenesis, specifically the time from a person's infection to symptom onset, leads to significant limitations on the geometries and statistical properties of time series at the population-level. Second, we address the evidence from outbreak investigation that a small fraction of infectious people are responsible for most of onward transmission, sometimes called the super spreader hypothesis \cite{endo2020estimating}, and we show that volatility at the population level can be used to dynamically assess this individual-level mechanism. Taken as a whole, we find that these mathematical relationships across epidemiological scales can be leveraged to create a single, stochastic transmission model capable of real-time inference and detailed situational assessment.

To be concrete, we develop these relationships by working with COVID-19 data from Washington state between January 2020 and March 2021. In other words, this paper focuses on roughly the pandemic's first year, before significant vaccination and before the emergence of variants of concern. This choice clearly limits the direct applicability of any of our results to today's decision-making, but we think that this emphasis on the early days helps to focus the discussion.

Speaking very generally before diving in, this paper's overarching theme is that the statistical mechanics, that is, the mathematical connections between mechanisms across scales, complements our biological understanding at each scale. In developing a couple of these connections, we end up creating tools that can be used to broadly characterize SARS-CoV-2 transmission and COVID-19 disease. And, as we demonstrate, quantities of public health interest like population prevalence, the lifetime of transmission chains, and the number of cases clustered as outbreaks can all be calculated naturally, with lightweight algorithms, in a single epidemiological framework.

\section{Pathogenesis as a building block}

To get started, we have to define ``situational awareness" with more mathematical precision. A concise, working definition is estimating $N_t$, the number of new COVID-19 infections every day, since many downstream quantities like reported cases or hospital admissions can be thought of as independent processes conditional on $N_t$. In other words, if we knew how many people were infected on a given day, and we assume that those people experience the disease independently once they have it, we can imagine them all flipping coins on their own to dictate their outcomes (recovery, death, etc.). 

One particularly important example of an independent process conditional on $N_t$ is the transition to infectiousness. SARS-CoV-2 is a respiratory pathogen, and the people who make up $N_t$ eventually become infectious after some time as the virus replicates in their bodies. This observation motivates defining an exposed but not yet infectious population, $E_t$, and an infectious population, $I_t$, such that
\begin{align}
        \hat{E}_t &= \left(1 - \frac{1}{d_E}\right)\hat{E}_{t-1} + \hat{N}_{t-1},\label{eq:e_t} \\
        \hat{I}_t &= \left(1 - \frac{1}{d_I}\right)\hat{I}_{t-1} + \frac{1}{d_E}\hat{E}_{t-1,}\label{eq:i_t} 
\end{align}
where hats mark expected values and $d_E$ and $d_I$ are the expected durations of the latent and infectious states. These equations establish a dynamic relationship between $\hat{E}_t$, $\hat{I}_t$, and $\hat{N}_t$, where new infections are a daily source term for each compartment's total population.

\begin{figure*}
\centering\includegraphics[width=\linewidth]{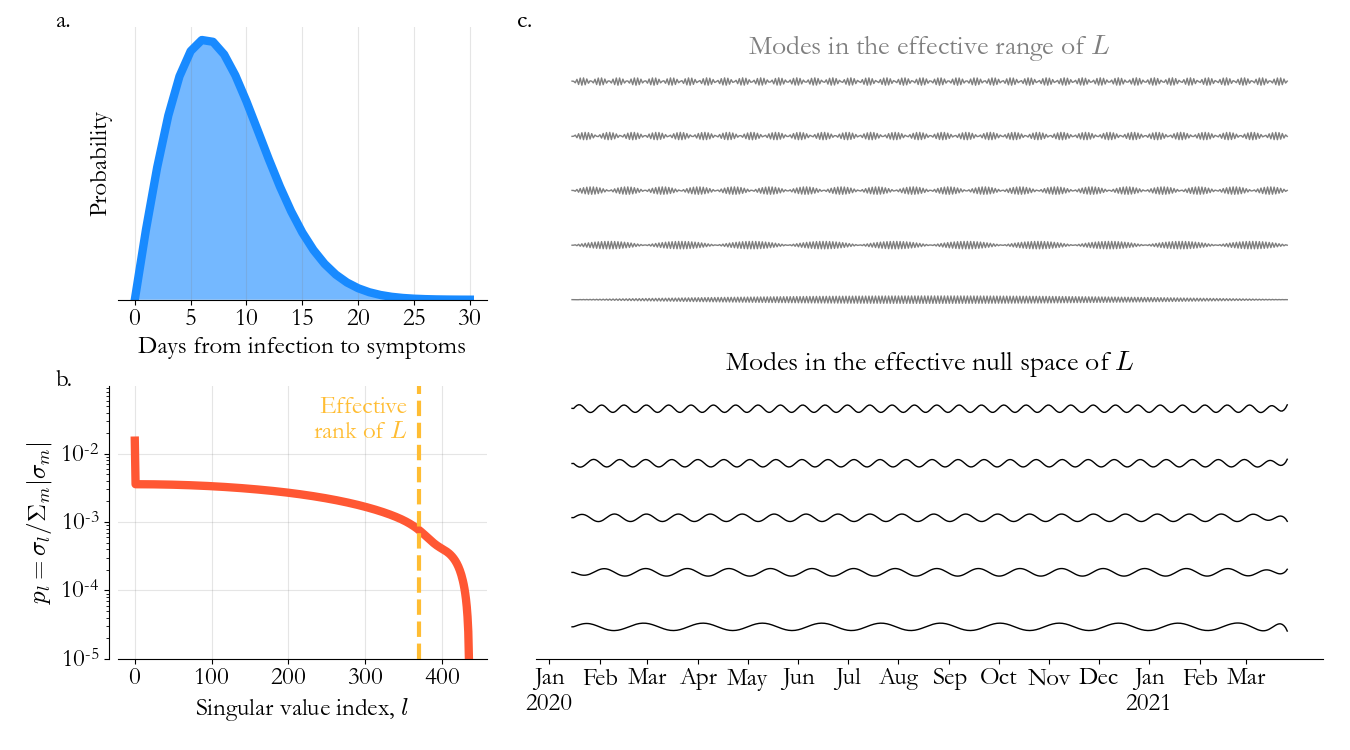}
\caption{The pathogenesis basis. (a) The distribution of the time from COVID-19 infection to symptom onset, characterizing individual-level pathogenesis, estimated in Ref.\ \onlinecite{qin2020estimation} using data on travel from Wuhan. (b) The singular value distribution (red) of $\mathbf{L}$, defined in Eq.\ \ref{eq:L}, with the threshold between statistically significant and insignificant modes (i.e. the null-space and range) marked in yellow. (c) Representative modes in $\mathbf{L}$'s range (noise) and null-space (signal) illustrating the implied smoothness of population-level time series consistent with (a).}
\label{fig:pathogenesis_modes}
\end{figure*}

The times $d_E$ and $d_I$ are often a major point of connection between the population and individual levels for any transmission model \cite{anderson1992infectious}. In practice, they're estimated indirectly from studies of disease progression, so-called pathogenesis studies, among closely observed individuals. A typical measurable input is the distribution of the time from exposure to symptom onset, $\pi(\tau)$, and one such estimate, by Qin \textit{et al.} \cite{qin2020estimation}, based on travel records of infected individuals leaving Wuhan, is visualized in Fig.\ \ref{fig:pathogenesis_modes}a. Specifying Eqs.\ \ref{eq:e_t} and \ref{eq:i_t} requires us to choose a $d_E$ and $d_I$ consistent with $\pi(\tau)$.

As mentioned, our goal is to deepen these types of connections. We can do so here by linking symptom progression to infectiousness under the assumption that symptom onset marks the mid-point of the infectious period. In that case, the pathogenesis distribution in Fig.\ \ref{fig:pathogenesis_modes}a is the expected time to infectiousness for an individual who just got COVID-19. Mathematically, writing $\tau=t-s$ for infection time $s$ and aggregating the expectation over newly infected individuals, this implies that
\begin{align}
    \hat{I}_t &= d_I\sum_{s=1}^T \pi(t-s)\hat{N}_s \implies \hat{\bm{I}}_t = \mathbf{P}\hat{\bm{N}}_t,\label{eq:gfunc}
\end{align}
where the time period of interest lasts $T$ days and bold-face is used to denote matrices and column vectors. The $T\times (T-2)$ matrix $\mathbf{P}$ has entries $P_{ts}=d_I\pi(t-s)$ and is 0 for $t<s$ since infection has to happen before infectiousness. This result is complementary to the more familiar Eqs.\ \ref{eq:e_t} and \ref{eq:i_t}, in the sense that it maps $\hat{N}_t$ to $\hat{I}_t$ through a global aggregating operation instead of a local differencing operation, somewhat like the connection between a Green's function and a linear differential equation. 

Inspired by that thought, if we rearrange Eqs.\ \ref{eq:e_t} and \ref{eq:i_t} to write 
\begin{align*}
    \hat{N}_{t-1} &= \hat{E}_t - \left(1 - \frac{1}{d_E}\right)\hat{E}_{t-1}\implies \hat{\bm{N}}_t = \mathbf{D}_E \hat{\bm{E}}_{t}, \\
    \hat{E}_{t-1} &= d_E\left[\hat{I}_t-\left(1 - \frac{1}{d_I}\right)\hat{I}_{t-1}\right]\implies \hat{\bm{E}}_t = \mathbf{D}_I \hat{\bm{I}}_t,
\end{align*}
defining weighted differencing matrices $\mathbf{D}_E$ and $\mathbf{D}_I$, we can relate $\hat{\bm{I}}_t$'s curvature to daily infections as $\hat{\bm{N}}_t = \mathbf{D}_E\mathbf{D}_I\hat{\bm{I}}_t$. Combined with Eq.\ \ref{eq:gfunc}, we find
\begin{align}
    \left(\mathbf{P}\mathbf{D}_E\mathbf{D}_I - \mathbb{I}\right)\hat{\bm{I}}_t = 0,\label{eq:L}
\end{align}
where $\mathbb{I}$ is the identity matrix. 

At first this may look like a boring relationship: If we take the appropriate differences of $\hat{\bm{I}}_t$ and then aggregate with $\mathbf{P}$, we get $\hat{\bm{I}}_t$ back. But for a given pathogenesis distribution and consistent latent and infectious durations, only some time series satisfy this equation. Intuitively, we've found that if a time series is constructed with collections of $\pi(\tau)$, then it must have certain geometric properties since arbitrary signals cannot be constructed with only that building block.

Completing this characterization of $\hat{\bm{I}}_t$ as an emergent outcome of pathogenesis requires us to analyze the linear operator $\mathbf{L}\equiv\mathbf{P}\mathbf{D}_E\mathbf{D}_I - \mathbb{I}$. In particular, we want to define the types of signals that $\mathbf{L}$ maps to values statistically indistinguishable from zero. Or, in equivalent algebraic terms, we need a basis for $\mathbf{L}$'s effective null space.

Fortunately for us, this is a classic problem in linear algebra and information theory, and there are nice theorems we can lean on. We recommend Ref.\ \onlinecite{roy2007effective} for specific details, but at a high-level, we can use the singular value decomposition to partition $\mathbf{L}$'s domain into an effective range and an effective null-space. The separation between the two can be estimated by computing the exponential entropy of the singular value distribution, which measures $\mathbf{L}$'s effective rank.\footnote{A little more detail here: The exponential entropy converges to the standard matrix rank if the non-zero singular values are uniformly distributed. In the more general case, you can construct a symmetric-positive-definite matrix for any singular value distribution (by squaring and square-rooting the matrix of interest), and if we interpret that matrix as the covariance matrix of a Gaussian process, the exponential entropy characterizes the number of uncorrelated degrees of freedom. That's essentially the same problem we have here, anticipating characterizing $I_t$ in terms of a mean and variance only.}

To be more concrete, we assume $d_E=5$ days and $d_I=4$ days (which we assume throughout the paper, see Ref.\ \onlinecite{thakkar2020towards} for details) and apply this approach to the pathogenesis distribution in Fig.\ \ref{fig:pathogenesis_modes}a. In Fig.\ \ref{fig:pathogenesis_modes}b, $\mathbf{L}$'s singular value distribution is plotted in red, with the boundary between the effective range and null-space, that is the information theoretic threshold for singular values that are effectively 0, overlaid in yellow. We see that the domain of possible $\hat{\bm{I}}_t$ time series from January 15, 2020 to March 26, 2021 has been condensed dramatically, from 437 independent dimensions to 66, roughly by a factor of the expected time to symptom onset.

In Fig. \ref{fig:pathogenesis_modes}c, representative modes confirm our previous intuition. Time series in the effective range have high-frequency and are inconsistent with $\pi(\tau)$. Meanwhile, time series in the null-space are considerably smoother and vary at scales slower than the time from infection to infectiousness. In effect, by concisely defining individuals in terms of their independent samples from $\pi(\tau)$, we've created a basis for population-level time series tailored to observed, COVID-19 specific pathogenesis.

\section{Signal processing in the pathogenesis basis}
Of course, we don't actually get to measure $\hat{I}_t$, and so it's not obvious how to put any of this to use. We can take a small step forward by noting that Eq. \ref{eq:L}, and the resulting modes, apply to any signal directly proportional to $\hat{I}_t$. So, as we approach the data from Washington, a reasonable goal, towards the broader goal of estimating $N_t$, is constructing a curve $\varphi_t \propto I_t$, which we'll refer to as the epi-curve.

Of the readily available data in Washington, daily cases, $C_t$, and hospital admissions, $H_t$, are likely the most closely related to $I_t$ since it's reasonable to assume both time series are a sample of individuals starting to experience symptoms, some time within their infectious period. More specifically, we model
\begin{align}
    C_t &= \alpha_t I_t + w_t \label{eq:Ct} \\
    H_t &= \beta_t I_t + v_t, \label{eq:Ht}
\end{align}
with $\alpha_t < 1$ and $\beta_t < 1$ representing $C_t$ and $H_t$ as dynamic subsets of the infectious population, sampled with observation noise processes $w_t$ and $v_t$. Since both $\alpha_t$ and $\beta_t$ change in time, neither $C_t$ nor $H_t$ are good candidates for the epi-curve on their own. 

\begin{figure*}
\centering\includegraphics[width=\linewidth]{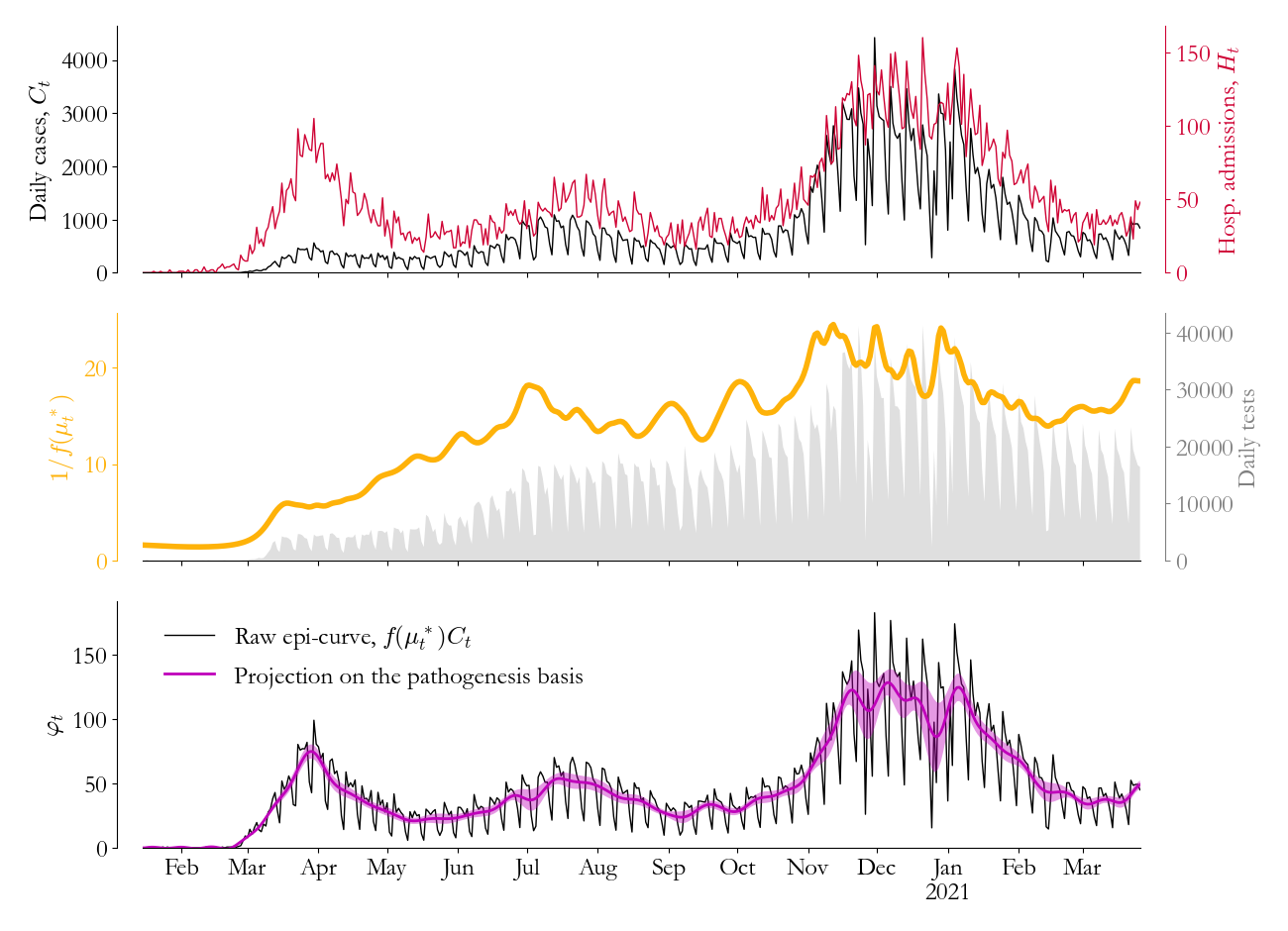}
\caption{Signal processing in Washington. (top) Daily case (black) and hospitalization (red) data show large differences in the relative sizes of the waves. (middle) The estimated, age-adjusted rolling fraction (yellow) of cases that are hospitalized, $f(\mu_t^*)$, is a measure of this discrepancy. We've plotted its inverse to highlight small changes. It is highly correlated with testing volume (grey), and it gives us information about the probability infections are reported as cases. (bottom) The raw epidemiological curve, $f(\mu_t^*)C_t$, (black) can be projected onto the pathogenesis basis to separate epidemiologically relevant signal (purple) from observation noise while dynamically quantifying volatility in transmission (95$\%$ CI shaded).}
\label{fig:pathogenesis_gpr}
\end{figure*}

This issue is illustrated in Fig.\ \ref{fig:pathogenesis_gpr}'s top panel. While $C_t$ (black) and $H_t$ (red) share common features, like the 3 distinct waves, they are certainly not scaled versions of the same signal, since they disagree on the relative wave sizes for example. Factors like the accessibility of testing and the age distribution of the infectious population must modify the relationships between $C_t$, $H_t$, and $I_t$. 

With those factors in mind, we can make some qualitative progress. Specifically, we know $\alpha_t \geq \beta_t$ for all time, since in general more people test positive than are admitted to the hospital. Second, we can think of $\alpha_t$ as a mixture of fast fluctuations incompatible with $\pi(\tau)$, like the apparent weekend effects in Fig.\ \ref{fig:pathogenesis_gpr}, and slow, systemic changes associated with testing infrastructure and demand. Meanwhile, we might expect fluctuations in $\beta_t$ to be smaller than in $\alpha_t$, corresponding to transient shifts in the average infection age, and we can model $\beta_t = \beta + \tilde{\beta}_t$ with $|\tilde{\beta}_t/\beta | < 1$.   

Under these assumptions, solving Eqs.\ \ref{eq:Ct} and \ref{eq:Ht} for $I_t$, setting them equal, and isolating $H_t$ to first order in $\tilde{\beta}_t/\beta$, we find
\begin{align*}
    \left(1 - \frac{\tilde{\beta}_t}{\beta}\right)H_t \approx \frac{\beta}{\alpha_t}C_t - \frac{\beta}{\alpha_t}w_t + \left(1 - \frac{\tilde{\beta}_t}{\beta}\right)v_t.
\end{align*}
This is very nearly a relationship we can use to estimate $\beta C_t/\alpha_t$, a good step towards $\varphi_t$ since its mean is directly proportional to $\hat{I}_t$. To get there, we need to estimate $\tilde{\beta}_t/\beta$, the fluctuations in the relative likelihood of hospitalization, and we need a tractable model for $\beta/\alpha_t$.

For the first problem, we refer to appendix A for details. Briefly, we leverage literature on the likelihood of severe COVID-19 infection as a function of age, specifically Ref.\ \onlinecite{verity2020estimates}. Incorporating the age distribution of the population from census and the observed age distribution of cases, we can estimate a dynamic probability of hospitalization given infection, and we can calculate the relative fluctuations around the mean $\tilde{\beta}_t/\beta$.

For the second problem, we recognize that $\beta/\alpha_t \leq 1$, inspiring us to write $\beta/\alpha_t = f(\mu_t)$ where $\mu_t$ is a Gaussian process (see appendix B) and $f(\cdot)$ is the logistic function. If we assume that $\mu_t$ varies on a 4 week timescale, isolating in $\alpha_t$ the slower changes in the health system and leaving for now the faster fluctuations incompatible with pathogenesis, we arrive at an approachable regression problem for $\mu_t$. Once that's solved, we can construct $f(\mu_t^*)$, the best estimate of $\beta/\alpha_t$, to compute a preliminary epi-curve estimate $f(\mu_t^*)C_t$.

The results of this approach are also plotted in Fig.\ \ref{fig:pathogenesis_gpr}. In the middle panel, $1/f(\mu_t^*)$ is visualized in yellow, showing that the relationship between $C_t$ and $H_t$ has moved through distinct phases associated with each wave. There are clear increases in March 2020, through the summer, and again in the fall, all associated with increased testing volume (shaded in grey), suggesting that $f(\mu_t^*)$ is behaving as expected. Meanwhile, in the lower panel, $f(\mu_t^*)C_t$ (black) illustrates the balance we've achieved: The epi-curve estimate looks at long time-scales like $H_t$ while retaining the short-time-scale fluctuations of $C_t$.

With this noisy estimate of $\varphi_t$ in hand, we're in a position to leverage the results in Fig.\ \ref{fig:pathogenesis_modes}. Specifically, we can model $N_t = \hat{N}_t + \epsilon_t/\beta$, introducing a transmission noise process $\epsilon_t$ scaled by $1/\beta$ for convenience in what follows. Then, $\mathbf{P}\bm{N}_t = \hat{\bm{I}}_t + \mathbf{P}\bm{\epsilon}_t/\beta$, and we can multiply Eq.\ \ref{eq:Ct} by $\beta/\alpha_t \approx f(\mu_t^*)$ to write
\begin{align}
    f(\bm{\mu}_t^*)\otimes\bm{C}_t = \hat{\bm{\varphi}}_t + \mathbf{P}\bm{\epsilon}_t + f(\bm{\mu}_t^*)\otimes\bm{w}_t,\label{eq:noise_corr}
\end{align}
where $\otimes$ denotes element-wise multiplication, and we've chosen $\varphi_t = \beta I_t$ so that $\hat{\bm{\varphi}}_t$ satisfies Eq.\ \ref{eq:L}. This result defines the two distinct sources of volatility. Transmission process noise, $\epsilon_t$, is the residual component with correlation mediated by $\mathbf{P}$ while the observation noise scales with $f(\mu_t^*)$ and remains independent day-to-day when $w_t$ is uncorrelated. In other words, a random fluctuation in the transmission process leads to observable changes smoothed in time by pathogenesis while volatility due to the measurement process evolves with the relationship between $C_t$ and $H_t$.

Eq. \ref{eq:noise_corr} motivates a signal processing approach to refine $f(\mu_t^*)C_t$ into its transmission-related components. Projecting $f(\mu_t^*)C_t$ onto $\mathbf{L}$'s effective null space to calculate $\hat{\varphi}_t$ gives us the residual $r_t \equiv f(\mu_t^*)C_t - \hat{\varphi}_t$. Then, we assume $\epsilon_t$ is a Gaussian process with correlation time $d_I$, modeling transmission fluctuations as driven by the behavior of infectious cohorts. Under that assumption, the joint probability distribution $p(\bm{\epsilon}_t,\bm{r}_t)$ is Gaussian as well, and we can compute\footnote{
There are some important practical details here, relegated to this footnote to help with more general reading comprehension. From Eq. \ref{eq:noise_corr}, $p(\bm{\epsilon}_t,\bm{r}_t)$ has a covariance matrix that depends on $\mathbf{W}$, the covariance matrix of $\bm{w}_t$. As as result, computing $p(\bm{\epsilon}_t | \bm{r}_t)$ requires us to specify $\mathbf{W}$.

Unfortunately, we can't just follow convention, assume $\mathbf{W} \propto \mathbb{I}$, and move on. The weekend suppression in testing is clearly a correlated observation effect. However, since we're only interested in $\epsilon_t$, which has longer time-correlation driven by $\mathbf{P}$, we can safely drop weekends and holidays from $r_t$, leaving days where we can more reasonably assume that $\mathbf{W} \propto \mathbb{I}$. Basically, because transmission on weekends has observable effects on weekdays, we can restrict our attention to days where $f(\mu_t^*)$ is a good estimate of the observation noise's relative scale. Practically, this means dropping the associated rows from $\bm{r}_t$ and $\mathbf{P}$ before making the usual assumption that $\mathbf{W} \propto \mathbb{I}$ to fully specify $p(\bm{\epsilon}_t|\bm{r}_t)$.} both $p(\bm{\epsilon}_t | \bm{r}_t)$ and the variance $\text{V}[\bm{\varphi}_t]=\text{E}[\mathbf{P}\bm{\epsilon}_t|\bm{r}_t]^2$ with standard methods \cite{bishop2006pattern}.
 
The results of this combined spectral/Gaussian process approach are visualized in purple in Fig. \ref{fig:pathogenesis_gpr}'s lower panel. The projection onto the pathogenesis basis extracts fluctuations in $f(\mu_t^*)C_t$ that are geometrically compatible with $\pi(\tau)$, naturally disregarding weekend and other fast fluctuations. Meanwhile, variance around $\hat{\varphi}_t$ changes dynamically, with time-correlation also defined by $\pi(\tau)$. Taken as a whole, developing an understanding of the population-level effects of individual-level pathogenesis has led us to an efficient, heteroskedastic signal-processing algorithm for COVID-19 time series.

\section{A mechanistic transmission model}

We're in a good place now with respect to our original goal, that is estimating $N_t$ as a step towards broader situational awareness. In fact, with $\varphi_t$ in hand, we can compute $\mathbf{D}_E\mathbf{D}_I\bm{\varphi}_t \propto \bm{N}_t$, so we're actually only one proportionality constant away.


Maybe unsurprisingly, there are a lot of ways we can anchor our estimates, especially in settings like Washington where we have survey data on COVID-19 prevalence. But a simple, general choice is through comparison with daily mortality, $M_t$, since the details of this outcome are well studied, for obvious reasons. 

To compute $\hat{m}_t \propto M_t$ given $\mathbf{D}_E\mathbf{D}_I\bm{\varphi}_t$, we leverage literature \cite{cdcifr2020} on the distribution of the time from exposure to death and the probability of death given infection (see appendix A). The calculation then proceeds along the lines of Eq.\ \ref{eq:gfunc}, and we can set the proportionality constant to the least-squares estimate, $\hat{\bm{m}}_t^T \bm{M}_t / \hat{\bm{m}}_t^T\hat{\bm{m}}_t$.

This scaling moves us from $\varphi_t$ to more complete estimates of $I_t$, $E_t$, and $N_t$, which as it stands are entirely descriptive, based only on features of the time series data as viewed through a particular lens. To then step into a population-level model, it's useful to relate these estimates to one another in a more mechanistic way. 

One option, following classic work in epidemiology \cite{anderson1992infectious}, is to note that new exposures come from an infectious population's interaction with a susceptible population. Writing $S_t$ for the susceptible population, this motivates a model where
\begin{align}
    N_t = \beta_t S_t I_t \varepsilon_t,\label{eq:seir}
\end{align}
introducing the average transmission rate $\beta_t$ and log-normal volatility $\varepsilon_t$, which jointly characterize the distribution of the fraction of susceptible-infectious pairs contributing to onward transmission every day. Then, if we assume that the total population is fixed,
\begin{align}
    S_t = S_{t-1} - N_{t-1} - V_{t-1},\label{eq:s_t}
\end{align}
which is a linear transformation of $N_t$ given an estimate of vaccine-derived immunity, $V_t$ (which we include for completeness, despite its negligible effect in the pandemic's first year, see appendix C). As a result, we can compute the mean and variance of $S_t$ from $N_t$ directly, assuming the population is fully susceptible at $t=0$. After that, we can estimate $\beta_t$ and $\text{V}[\varepsilon_t]$ by approximating the moments $\text{E}[N_t/I_tS_t]$ and $\text{V}[N_t/I_tS_t]$ in a first order Taylor series around the mean.

With the previous section's signal processing distilled into $\beta_t\varepsilon_t$, Eqs. \ref{eq:seir} and \ref{eq:s_t} can be combined with Eqs. \ref{eq:e_t} and \ref{eq:i_t} (removing the hats) to complete a mechanistic model of SARS-CoV-2 transmission, at least up to initial conditions. For our purposes in Washington, to resolve this last practical detail, we assume that $t=0$ corresponds to January 15, 2020, but that the time series data is reliable starting March 1, 2020. For the time in between, we assume $E_0 = 0$ and that prevalence grows exponentially, and we set $I_0$'s mean and variance to guarantee growth into the associated $I_t$ estimates on March 1.\footnote{This is basically an arbitrary choice, but without data during this time period, a lot of choices work. We picked this method only because it's simple, and we try to be content with a model that can't speak very specifically about this month and a half.} 

\begin{figure*}
\centering\includegraphics[width=\linewidth]{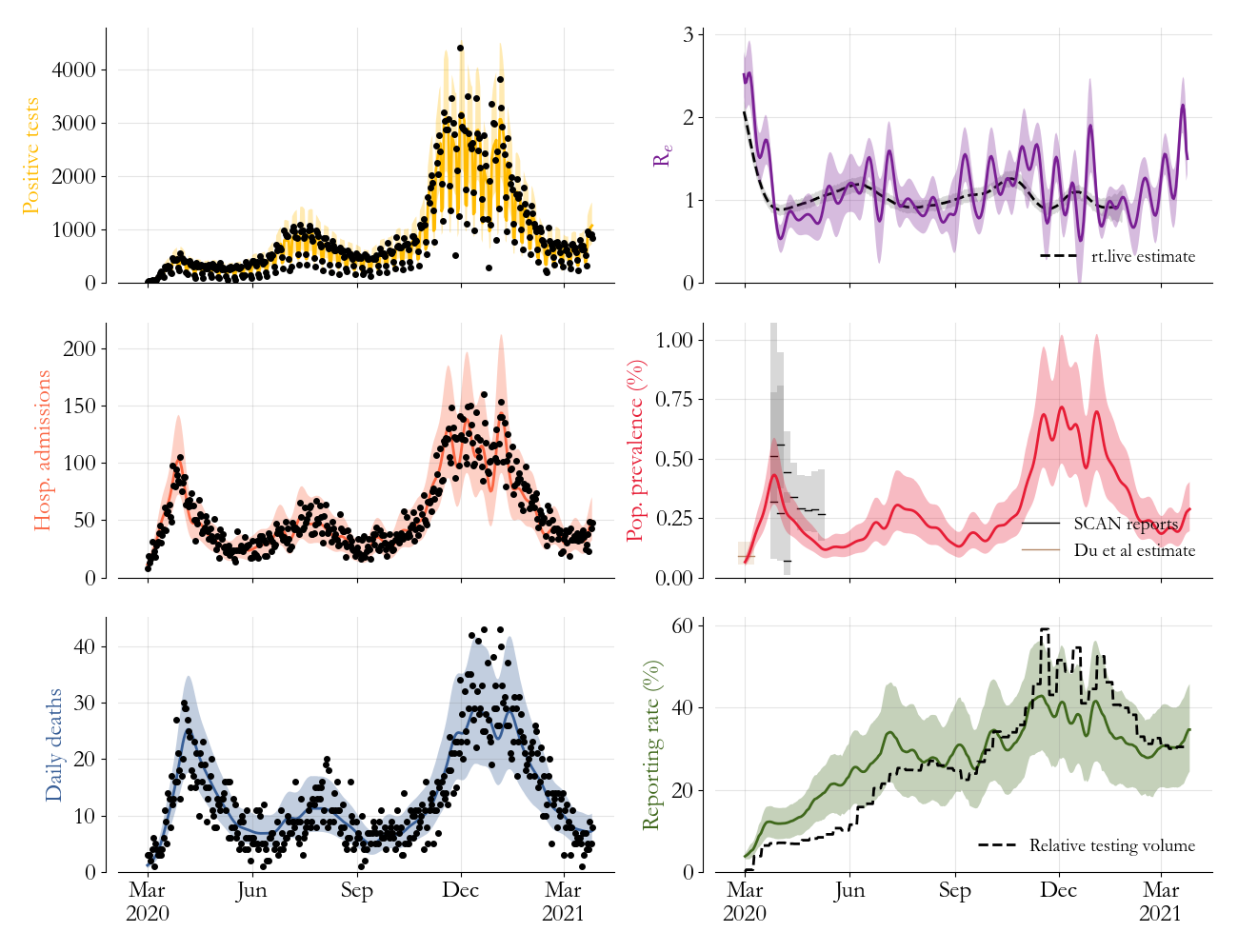}
\caption{A compartmental, stochastic process model of SARS-CoV-2 transmission in Washington. Adding mortality data to cases and hospitalizations (black dots) gives us enough information to fully specify a transmission model (colors, 95$\%$ CI shaded). The model trajectories (10,000 samples) are consistent with all 3 observed time series (left panels). The same trajectories can be used to estimate quantities of interest like $\text{R}_{e}$ (purple), COVID-19 prevalence (red), and the probability that infections are reported as cases (green).}
\label{fig:rainier}
\end{figure*}

Now, with the transmission model fully specified, we can assess its quality by sampling a bunch of trajectories and using them to compute things. We visualize 6 illustrative examples in Fig.\ \ref{fig:rainier}.

In the figure's first column, we sample observable outcomes across the ensemble. This requires us to model the observation process, which we choose to do coarsely since we want to remain focused on transmission. Specifically, we assume
\begin{align*}
    M_t &\sim \text{Binomial} \left\{N_{t-\tau_M},p^M_{t-\tau_M} \right\}, \\
    H_t &\sim \text{Binomial} \left\{N_{t-\tau_H},\theta_0 p^H_{t-\tau_H} \right\}, \\
    C_t &\sim \text{Binomial} \left\{I_t, (\theta_1+\theta_2\mathbf{1}_{t,\text{w}})/f(\mu_t^*) \right\},
\end{align*}
where $\tau_M$ and $\tau_H$ are the times from exposure to the associated outcome and $p_t^M$ and $p_t^H$ are the outcome probabilities mentioned previously, the latter of which we scale by a global factor, $\theta_0$, to account for Washington-specific health-seeking behavior. Meanwhile, for reported cases, the daily detection rate $\alpha_t = (\theta_1 + \theta_2\mathbf{1}_{t,\text{w}})/f(\mu_t^*)$, requires us to estimate another global scale factor $\theta_1$ and a weekend correction $\theta_2\mathbf{1}_{t,\text{w}}$. All three factors, $\theta_0$, $\theta_1$, and $\theta_2$ are estimated via least squares between the model average and the data. By construction, the model trajectories (colors) maintain consistency with the data (black dots). But critically, a single set of trajectories is able to simultaneously reconcile the superficially different patterns in cases, hospitalizations, and deaths.

In Fig. \ref{fig:rainier}'s second column, we turn to quantities of interest that are difficult to measure directly. In the top panel, $\beta_t\varepsilon_t$ is used to calculate the effective reproductive number, $\text{R}_e = d_IS_0\beta_t\varepsilon_t$. The resulting estimate (purple) is broadly comparable to other published \cite{rtlive2020} estimates (black), showing the effects of mitigation efforts early on but retaining fluctuations consistent with the time-scales encoded in the pathogenesis distribution. 

In the next panel, we calculate $E_t + I_t$ across the ensemble to estimate prevalence, showing clearly the shape retained from $\varphi_t$ in Fig.\ \ref{fig:pathogenesis_gpr}. Overlaid are multiple estimates from Seattle-focused prevalence surveys \cite{scanreport1,scanreport2,du2020using} during the first wave, showing good agreement in general.

Finally, in the last panel, we compute the weekly reporting rate over the ensemble, by comparing $C_t$ to the $I_t$ trajectories in a 7-day rolling average, accounting for the fact that each infection can only be detected once. We see the that the mechanistic model retains the correlation with testing volume (black) from Fig.\ \ref{fig:pathogenesis_gpr}, but now, with a more interpretable vertical axis, the estimate suggests diminishing returns of volume increases and potential disruptions during the holiday season.

Take as a whole, Fig.\ \ref{fig:rainier} demonstrates that the model provides a versatile, population-level picture of SARS-CoV-2 transmission in Washington state, capable of capturing observations with a biologically reasonable underlying dynamic. Moreover, maybe somewhat surprisingly, this nonlinear model was constructed through combinations of efficient linear operations, like the SVD and Gaussian process regression. As a result, on a laptop, moving from raw data to Fig.\ \ref{fig:rainier} takes only a few seconds.

\section{The branching process perspective}
One striking feature of the previous section's model construction is that we never need to make assumptions about the underlying social structure. Because $\beta_t\varepsilon_t$ is in principle free to vary in Eq.\ \ref{eq:seir}, we only need to assert that individuals can be meaningfully classified as susceptible or infectious and that pair-wise interactions between these two groups lead to new exposures. 

That said, we know infectious disease transmission is a social process. And so it's natural to ask: To what extent can we use the model to characterize the person-to-person interactions driving transmission? 

We can look to classical branching process theory \cite{fewster2014stochastic} for inspiration on this question. A necessarily equivalent, alternative formulation of Eq.\ \ref{eq:seir} is
\begin{align}
    N_t = \sum_{i=1}^{I_t} T_{it}, \label{eq:bp}
\end{align}
where $T_{it}$ is the number of people infected by the $i$th infectious individual on day $t$. Unlike Eq.\ \ref{eq:seir}, which describes transmission as a fraction of potential interactions, this equation describes the same process as a realized total of transmission events, emphasizing an individual-level perspective. 

The structure in Eq.\ \ref{eq:bp} is often called a ``random sum of random variables", since both the number of terms in the sum and terms themselves are uncertain. These types of probabilistic structures are well-studied and somewhat intuitively, the statistics of $N_t$, $I_t$, and $T_{it}$ are all inter-dependent. 

More specifically, if we define $T_{it}$ on a given day as an independent realization of a random variable $T_t$, the laws of total expectation and variance imply that
\begin{align*}
    \text{E}[T_t] &= \text{E}[N_t]/\text{E}[I_t]\\
    \text{V}[T_t] &= (\text{V}[N_t]-\text{E}[T_t]^2\text{V}[I_t])/\text{E}[I_t],
\end{align*}
relating $T_t$'s mean and variance to those of $N_t$ and $I_t$ directly. In other words, we can characterize $p(T_t)$, the sampling distribution of infectious people's epidemiologically relevant social contacts on day $t$, through statistics computed across the same bundle of trajectories used to make Fig.\ \ref{fig:rainier}.

\begin{figure*}
\centering\includegraphics[width=0.98\linewidth]{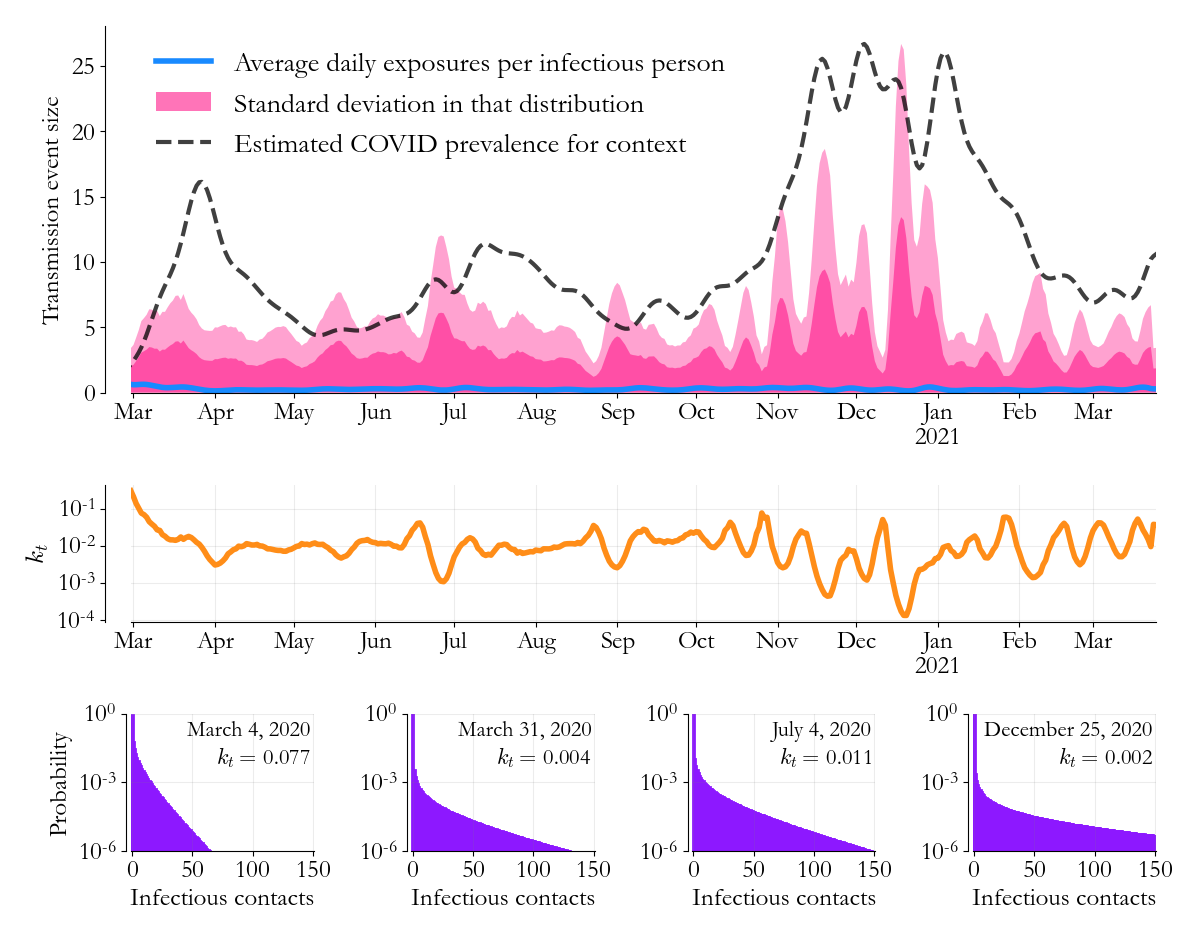}
\caption{Person-to-person transmission in the model. The daily number of transmissions per infectious individual has low mean (blue) and large variance (1 and 2 standard deviations in pink tints). The volatility is dynamic, and periods of high volatility are associated with prevalence increases (black dashed line). Matching negative binomial distributions to the mean and variance gives us a dynamic characterization of over-dispersion relative to a Poisson distribution ($k_t$, in orange) and a detailed, probabilistic picture of infectious contacts (purple) consistent with the time series data in Washington.}
\label{fig:overdispersion}
\end{figure*}

The results of this calculation are visualized in Fig.\ \ref{fig:overdispersion}. In blue, we see that the average daily number of transmissions per infectious individual, $\text{E}[T_t]$, is relatively low, significantly below 1. Meanwhile, standard deviations around that average (1 and 2 deviations are shaded in pink) grow to much larger numbers, sometimes as high as 25. In qualitative agreement with findings from contact tracing \cite{sun2021transmission}, we find that the individual-level infectious contact distributions are highly skewed. But here, we can resolve this skewness in time, and by overlaying the average prevalence from Fig.\ \ref{fig:rainier} (black), we also find that high volatility in $T_t$ is associated with epidemic growth while volatility suppression is associated with control.

With a little more effort, we can add a lot of quantitative detail to our understanding of these person-to-person interactions. Specifically, inspired by classical statistical mechanics and information theory \cite{jaynes1957information}, we can search for an entropy maximizing distribution consistent with $E[T_t]$ and $V[T_t]$, given that $T_t$ can only take non-negative integer values. Solving this inference problem would give us estimates of the full distributions $p(T_t)$ for every day.

Unfortunately, this problem is analytically intractable in general, but there are some nice theorems we can use to find an approximate solution. Specifically, from Ref.\ \onlinecite{harremoes2001binomial}, we know that the binomial distribution is the entropy maximizing distribution of a sum of independent but non-identical Bernoulli trials with fixed mean. That's essentially our situation, imagining a single infectious individual in a fully connected, weighted network, with each connection an independent but non-identical (and potentially 0 success probability) coin-flip for a new infection. 

If we then imagine that the fixed mean is instead Gamma distributed, and we take the limit of a large network with low connection weights, we get a Gamma-Poisson mixture, equivalent to the negative binomial distribution. Lucky for us, that's the choice we probably would've made based on convenience alone, but it's reassuring to know that we have a little theoretical footing. 

In any case, we can analytically match negative binomial distributions to $\text{E}[T_t]$ and $\text{V}[T_t]$ for all $t$, giving us
\begin{align*}
    p(T_t = \ell) = \frac{\Gamma(\ell+k_t)}{\Gamma(\ell+1)\Gamma(k_t)}\left(\frac{\mu_t}{\mu_t+k_t}\right)^{\ell}\left(\frac{k_t}{\mu_t+k_t}\right)^{k_t},
\end{align*}
where $\Gamma(\cdot)$ is the gamma function, $\mu_t = \text{E}[T_t]$, and $k_t = \text{E}[T_t]^2/(\text{V}[T_t] - \text{E}[T_t])$. By construction, these contact distributions reproduce the summary statistics in Fig.\ \ref{fig:overdispersion}, which as a consequence of Eq.\ \ref{eq:bp}, further reproduce the results in Fig.\ \ref{fig:rainier}, thus maintaining consistency with the observed time series in Washington.

We visualize $k_t$ (orange) and some representative infectious contact distributions in Fig. \ref{fig:overdispersion}'s bottom half. In general, all the distributions are heavy tailed, and averaged over time, $96\%$ of infectious individuals infect no one on a given day, implying that $85\%$ infect no one at all during the course of their infection, in good agreement with contact tracing in other settings \cite{sun2021transmission}. That said, early in the model period, on March 4, 2020, that daily number is considerably lower, $84\%$. By the end of the month, as prevalence grows and as mitigation efforts are implemented, $p(T_t = 0)$ grows as well.

These sampling distributions for $T_t$ clearly support a super spreader mechanism for SARS-CoV-2 transmission in Washington, with a small fraction of infectious individuals responsible for all of a given day's transmission. That we calculated these distributions from only population-level time series tells us that, when properly viewed, the population-level data contains signatures of this individual-level heterogeneity.

\section{Characterizing the transmission forest}
These inferred contact distributions give us an opportunity to build connections to other individual-level measures of transmission, like outbreak investigation and phylogenetics. In building these connections, we can also look to further validate the results in Figs.\ \ref{fig:rainier} and \ref{fig:overdispersion}. 

In Washington, public health officials generally define an outbreak as a collection of two or more cases with evidence of concurrent exposure in a shared location that isn't a household. We note though that this definition gets some nuance in health-care settings, like in long-term care facilities for example, where either a single case in a patient or two or more cases in workers counts. Overall, these outbreak reports are aggregated into weekly time series by an approximate outbreak start date and then regularly published in documents like Ref.\ \onlinecite{wadoh2021}.

We can coarsely compare our estimates to these time series by approximating the outbreak reporting probability as a collection of three independent events: that an outbreak happens, that two or more associated infections test positive, and that those individuals are interviewed by contact tracers.

More specifically, we model the probability of an outbreak on day $t$ as $p(T_t \geq 2)$. Then, leveraging our previous estimate of the probability that an infection is reported as a case (from Fig.\ \ref{fig:rainier}, in green), in combination with data on the weekly fraction of cases that contact tracers interview, we further estimate the probability that at least 2 cases from the outbreak are reported and investigated. Putting these together with the model's $I_t$ trajectories leads us to the number of $T_t \geq 2$ events investigated per week, which we discount by a global factor to account for the portion associated with household transmission.

\begin{figure*}
\centering\includegraphics[width=0.98\linewidth]{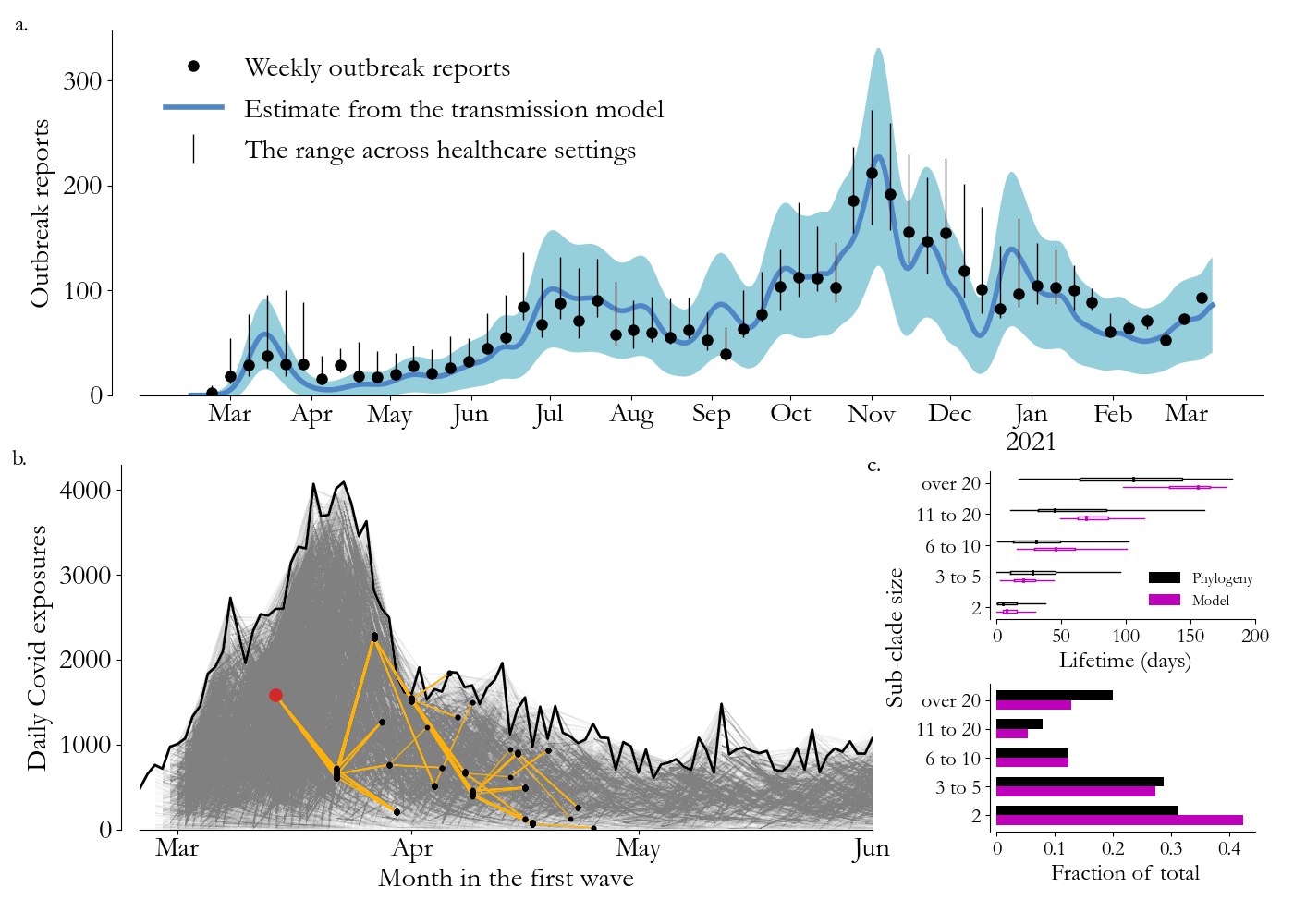}
\caption{Connections to outbreak investigation and phylogenetics. (a) Investigators in Washington search for outbreaks (generally 2 or more cases likely exposed at the same time and place) across community and healthcare settings (black dots). The model estimate (blue), based on the inferred contact distributions and adjusted for reporting, is consistent with their findings throughout the model period. (b) The contact distributions can also be used to sample approximate transmission trees (grey) that reproduce model trajectories (black) when summed. (c) Analysing individual transmission trees within this forest, again accounting for reporting, the model broadly reproduces statistics estimated with genetic sequencing data.} 
\label{fig:validation}
\end{figure*}

The results of this calculation are compared to the outbreak investigation reports in Fig.\ \ref{fig:validation}a. The model-based estimate captures key features of the data, like the steady rise into the summer and the distinct multi-peaked structure in the winter. Moreover, in making this comparison, we learn through the least-squares estimated scale-factor that roughly $60\%$ of Washington's $T_t \geq 2$ events happen in households, in reasonable agreement with targeted serological surveys in other settings \cite{bi2021insights}. 

This comparison is confidence building, at least when it comes to a broad statistic of $p(T_t)$. More generally, to make more granular connections, it's useful to be able to sample transmission trees consistent with a given model trajectory. Similar to the classical branching process, our contact distributions can be used to do that as well.

For a particular $N_t$ trajectory, note that
\begin{align*}
    \sum_{i=d_E +1}^{d_E+d_I} N_{t-i} \approx I_t,
\end{align*}
for $t \geq d_E + d_I$ since the infectious parents of day $t$'s new exposures were themselves new exposures some time $d_E$ to $d_E + d_I$ days earlier, approximating the infectious duration as deterministic and again assuming that the population is closed. If we round $N_t$ to the nearest integer, this equation defines two populations of nodes, infectious parents and their children, arranged in time along $N_t$ and linked in a bipartite graph. 

With this picture in mind, consistency with the transmission model implies that all children have only one parent and that the joint distribution of children per parent is $p(T_t | \sum_i^{I_t} T_{it} = N_t)$. Thus, moving through time, drawing one such graph per day (see appendix D), we can create sample transmission trees by linking graphs under the assumption that infectious individuals have no memory of their history. For a particular model trajectory, this approach generates a collection of independent trees, generally called a forest, each grown from the initial seeds corresponding to the $N_t$ with $t<d_E + d_I$. 

A sample transmission forest made in this way is visualized across Washington's first wave in Fig.\ \ref{fig:validation}b. Nodes are arranged in columns on their exposure date, emphasizing that their daily sum reproduces the $N_t$ trajectory (black line), and that as a result, outcomes computed across these trees live within the intervals in Fig.\ \ref{fig:rainier}. 

The forest is large, and difficult to take in, with well over 100,000 edges (grey), and so we also highlight a representative transmission chain in yellow. In that chain, the red node is a super spreader, infecting 122 people and leading to 315 more infections (black) before the chain's stochastic extinction in late April. The grey lines represent many such chains, all overlaid on top of one another in an intricate geometry consistent with $p(T_t)$.

The yellow chain's finite lifetime motivates a high-level comparison to Washington's genetic sequencing data, arranged into a phylogenetic tree in Ref. \onlinecite{tordoff2021phylogenetic}. In that paper, the phylogenetic tree is separated into subclades, groups of ancestrally related sequences arranged over time, associated with a single SARS-CoV-2 importation and subsequent transmission. The authors find an empirical relationship between the number of sequences in the subclade and the subclade's lifetime.

We can check that independent transmission trees within Fig.\ \ref{fig:validation}b's sample forest reproduce these statistics. Specifically, breaking the forest into its disconnected components, we can coarsely estimate a subclade size by sampling nodes to account for their probability of being reported (Fig.\ \ref{fig:rainier}, again) and their probability of being sequenced (roughly $6\%$ \cite{tordoff2021phylogenetic}). Then, we can also calculate a subclade lifetime from the exposure dates of the first and last sequenced nodes.

This comparison is visualized in Fig.\ \ref{fig:validation}c, with the phylogenetic estimates in black and the model estimates in purple. The sample forest in Fig.\ \ref{fig:validation}b has a population of transmission trees comparable to the phylogeny's subclades, with lifetimes proportional to size and with some trees lasting longer than 6 months. Meanwhile, the distribution of approximate subclade sizes is non-uniform, reproducing the observation from the phylogeny that small subclades are over represented.

The transmission forest in Fig.\ \ref{fig:validation}b is certainly an over-simplification, lacking for example the importations that would be needed to explain the variants of concern to come. But despite that, perhaps because importations are a minority of all events \cite{tordoff2021phylogenetic}, the contact distributions $p(T_t)$ seem to be generally consistent with other, more individualistic measures of the interactions driving transmission. That they can be calculated directly from the model trajectories, essentially by hand, is remarkable. 

\section{Conclusion}

The main idea of this paper is that developing concise, mathematical connections across epidemiological scales leads to an efficient situational awareness more or less organically. We developed 2 specific connections, encapsulated in Eqs.\ \ref{eq:L}, \ref{eq:noise_corr}, and \ref{eq:bp}, and we used them to reconcile a few measures of transmission within a single, relatively standard stochastic process model. 

In the literature, transmission models with this somewhat simplistic compartmental structure are often abandoned for more computationally complex approaches. There are sometimes appealing reasons to take that path, but we think that the results in this paper make a case for the epidemiological nuance that can be captured by more classically inspired methods.  

\section*{Acknowledgements}
This work was done with constructive input from many colleagues. In particular, we want to thank Josh Herbeck, Greg Hart, Rafael Nunez, Jen Schripsema, Harrison Goldwyn, and Greer Fowler for their insight and attention. We also want to thank our colleagues at the Washington Department of Health, particularly Ian Painter and Gita Singh, who both have incredible perspective on COVID-19 data collection and public health practice, among other things.

\appendix
\section{Hospitalization and fatality probabilities}
To relate $N_t$ to $H_t$ and $M_t$, we make use of published estimates of severe infection \cite{verity2020estimates} and mortality \cite{cdcifr2020} probabilities as a function of age. We use these to construct all-population averaged estimates that account for transient changes in the age distribution of the infectious population and for advances in COVID-19 treatment. For an in-depth discussion, see Ref.\ \onlinecite{thakkar2020one}.

At a high level, for a given outcome, $o$, the outcome probability $p(o|I) = \sum_a p(o | I, a) p(a |I)$, where $I$ marks COVID-19 infection and $a$ is a 10-year age bin. The age distribution, $p(a|I)$, is estimated in 2 ways: First, assuming that the infectious population has the same age distribution as the total population and second, assuming that the weekly age distribution of cases is representative. $p(o|I)$ is constructed under each approximation and then treated respectively as a prior and likelihood in a Bayesian update approach. The mean and variance of this overall estimate is then linearly interpolated from the weekly to the daily timescale.

Finally, for mortality in particular, we use the results of a separate survival analysis \cite{sashidhar2020survival} to adjust the infection-fatality probability for advances in treatment. Meanwhile, as mentioned in the main text, for hospitalization, the relationship between severe infection and hospital admission is estimated with an overall scale factor.

\section{Gaussian processes}
In Eq.\ \ref{eq:noise_corr}, we use Gaussian processes to model time correlation in $\bm{\mu}_t$ and $\bm{\epsilon}_t$. Mathematically, this means that we assume both vectors have multivariate Gaussian prior distributions with 0 mean and covariance matrices $\mathbf{\Sigma}_{\mu}$ and $\mathbf{\Sigma}_{\epsilon}$ respectively.

For $\bm{\epsilon}_t$, we define $\mathbf{\Sigma}_{\epsilon}$ in terms of an exponential correlation function \cite{bishop2006pattern}, $k(t,s) = A\exp(-(t-s)^2/2\tau^2)$, with $\tau = d_I$ to model correlated behavior across infectious cohorts. The overall level, $A$, is based on the empirical performance of the intervals in Fig.\ \ref{fig:rainier}'s left column. In sensitivity testing, the paper's results were quantitatively insensitive to reasonable changes in $A$, and we use $A=0.35$ throughout.

For $\bm{\mu}_t$, it's more computationally convenient to write $\mathbf{\Sigma}_{\mu}$ in terms of its inverse, the precision matrix $\mathbf{\Lambda}_{\mu}$. We choose $\mathbf{\Lambda}_{\mu} = \lambda \mathbf{D}^T\mathbf{D}$, where $\mathbf{D}$ is the finite-difference approximation to the second derivative. The constant $\lambda$ can be related to a correlation time $\tau$ by noting that the expected total variation in $\bm{\mu}_t$, $||\mathbf{D}\bm{\mu}_t||^2$, is inversely proportional to $\lambda$. Meanwhile, for a sine-wave with period $\tau$, the total variation goes as $\tau^{-4}$, implying that setting $\lambda \propto \tau^4$ gives signals a prior expected time-scale $\tau$. As mentioned in the main text, we choose $\tau = 4$ weeks in the regression problem for $\bm{\mu}_t$.

\section{Vaccination in the model}

COVID-19 vaccine was introduced in Washington in mid-December 2020. It has a negligible effect from this paper's perspective since we focus on the pandemic's first year. We include it in the model mainly for illustrative purposes, to show how other sources of immunity might be incorporated into overall susceptibility. 

Per dose vaccine efficacy is modeled as a Beta distribution, with shape parameters $a=158.7$ and $b = 114.7$, set to approximately recapitulate observational study results \cite{dagan2021bnt162b2}. Under that simplistic efficacy model, we use daily time series data on the number of first and second doses administered in Washington to estimate the mean and variance in the number of individuals getting successfully vaccinated, assuming that immunity steps up 21 days after the dose. 

Finally, to calculate the daily population immunized, $V_t$, we take this estimate of successful vaccinations and adjust by the fraction of the population with natural immunity. This discounts for the possibility that someone with prior immunity gets successfully vaccinated, under a coarse assumption that vaccination and COVID-19 risk are uncorrelated.

\section{Sampling graphs and trees}
For a particular day, given $I$ infectious nodes and new exposures $N$, since all $N$ have only 1 parent, drawing a bipartite graph is equivalent to writing a length $I$ vector of parent-child connections, $g$, that sums to $N$. To construct a vector in accordance with the associated contact distribution $p(T)$, we start by writing
\begin{align*}
    g = (1, 1, 1, ..., 1, 0, 0, 0, ..., 0),
\end{align*}
a vector of $N$ ones and $I-N$ zeros. Then, starting at $g$'s first entry ($i=0$) and defining the rolling sum $G = \sum_{j=0}^{i-1} g_j$, we draw a sample, $\ell$, from
\begin{align*}
    p(T|N,G) = \frac{p(T)}{1 - \sum_{T > N-G} p(T)},
\end{align*}
the conditionally renormalized degree distribution. We then set $g_i = \ell$, set the next $\ell-1$ entries to 0, and then repeat the process with $i\rightarrow i+\ell$ until $G = N$.

This rewiring approach scales linearly in $N$ and is, as a result, efficient enough to be performed every day for a given model trajectory, starting at $t=d_E+d_I$ so that both $I$ and $N$ are well defined. Daily graphs are then connected by assigning positions in $g$ to the nodes $I$ uniformly randomly, without regard to past edges already assigned to any node.

\bibliography{references}

\end{document}